\documentclass[a4paper,12pt]{article}
\usepackage{amsmath}
\usepackage{amsthm}
\usepackage{amssymb}

\title{On fixed and uncertain mixture prior weights}

\author{Beat Neuenschwander, Simon Wandel, Satrajit Roychoudhury, Heinz Schmidli}

\begin{document}
\maketitle

\date{}


\tableofcontents

\section{Introduction}
Prior distributions play an eminent role in Bayesian statistics. Here, we are concerned with mixture prior distributions, with a  focus on the specification of the weights for the components of  mixture priors.

Mixture prior distributions allow statisticians to extend the set of simple prior distributions. They arise in various situations, for example:
\begin{itemize}
\item 
Dallal and Hall \cite{dal1983apb}
and
Diaconis and Ylvisaker \cite{dia1984qpo} showed that 
mixture distributions can be used to approximate any given prior distribution to any degree of accuracy. This is particularly useful for non-standard prior distributions. For example, meta-analytic predictive prior distributions derived from historical data have recently been discussed by 
Neuenschwander et al. \cite{neu2010shi}, 
Spiegelhalter et al. \cite{spi2004bac}, 
and Viele et al. \cite{vie2014uhc}. 
These priors arise from a random-effects meta-analysis of historical data to predict the parameter in the new study. Since MCMC analyses are required, the meta-analytic prior distribution is only available as a Monte Carlo sample. Such distributions can easily be approximated by mixtures of standard prior distributions using software such as the Bayesian R-package \emph{RBesT} 
(Weber \cite{web2017rbest}) 
or \emph{SAS PROC FMM} 
\cite{SAS2014FMM}. 
\item 
The Bayesian inference based on an informative prior distribution can be robustified by extending the original prior with a second, weakly-informative mixture component. The modified prior is an example of a heavy-tailed prior, which allows for more robust inferences in case of prior-data conflict 
(O'Hagan \cite{oha2004bi}, 
O'Hagan and Pericchi \cite{oha2012bhm}).  
Recent applications of such priors in medical product development include 
Mutsvari et al. \cite{mut2016app}, 
Schmidli et al. \cite{sch2014rmp}, 
Walley et al. \cite{wal2015aba}, and 
Papanikos et al. \cite{pap2020bhm}.
\item
Mixture priors can be used to represent expert opinions
(Dallow et al. \cite{dal2018bdm}, Moatti et al. \cite{moa2013moe}) or prior scenarios, including optimistic and pessimistic priors   
(Gajewski and Mayo \cite{gaj2006bss}).
\end{itemize}

While the above examples are often based on a single mixture prior (for the main parameter of a statistical model), the following applications involve multiple mixture priors:
\begin{itemize}
  \item 
Variable selection in regression analysis can be represented by a two-component mixture prior for each regression coefficient: the first component is a point-mass or "spike" prior at $0$, whereas the second component is a "slab" prior away from $0$. Such applications arise in various fields; see, for example,   
Fridley et al. \cite{fri2010bmm},
Fr\"{u}hwirth-Schnatter and Wagner \cite{fru2010bvs},
Ley and Steel \cite{ley2008ote},
Malsiner-Walli and Wagner \cite{mal2011csa},
Vanucci and Stingo \cite{van2010bmf},
and Xu and Ghosh \cite{xu2015bvs}.
\item
Subgroup or subpopulation identification also uses "spike-and-slab" priors. For example,  
Berry and Berry \cite{ber2004afm} and  
Xia et al. \cite{xia2011bhm}
have used such priors to detect safety signals in clinical trials.
\end{itemize}

Section \ref{sec::methods} discusses fixed and uncertain weights for the cases of a single and multiple mixture priors. Finally, the findings are summarized and some recommendations are suggested in Section  
\ref{sec::discussion}. 

\section{Methods}
\label{sec::methods}

\subsection{One mixture prior}
\label{sec:mixprior1}

We first consider the case of a single k-component mixture prior distribution
for the parameter $\theta$ of a statistical model $f(Y \vert \theta)$, that is
\begin{equation}
\label{mixprior::fixed1}
f(\theta) = \sum_{i=1}^k p_i f_i(\theta), \quad \sum_{i=1}^k p_i=1
\end{equation}
The parameters of the prior $f(\theta)$ are the mixture weights $p_i$ and the parameters of the component distributions $f_i(\theta)$, which have been omitted from notation. 
An alternative to (\ref{mixprior::fixed1}) is the representation with a latent variable $Z$ taking values 
$1$ to $k$
\begin{equation}
\label{mixprior::fixed2}
f(Z=i) = p_i, \quad f(\theta \vert Z=i) = f_i(\theta) \quad i=1, \ldots k 
\end{equation}
The inference for $\theta$ is the same for both versions because the distribution of $Y$ depends only on $\theta$ and the marginal prior distribution of the augmented prior (\ref{mixprior::fixed2}), that is, 
$f(\theta) = \int_{Z} f(\theta \vert Z)f(Z)dZ$, 
which is equal to (\ref{mixprior::fixed1}). However, the latent variable version has advantages: posterior mixture weights $pr(Z = i \vert Y)$ can be obtained (Section \ref{weights::fixed}), and Markov Chain Monte Carlo computations via Gibbs sampling can often be done easily (Appendix \ref{sec::appex1}).
Moreover, the latent variable representation will be critical for a better understanding of fixed and uncertain weights.

We now discuss fixed and uncertain weights. 
Two concerns against fixed weights are sometimes raised:
\begin{enumerate}
  \item that they prohibit dynamic updating
  \item that they may not reflect potential uncertainties of the weights.
\end{enumerate}
We will show that both concerns are unwarranted.

\subsubsection{Fixed mixture weights}
\label{weights::fixed}
For fixed weights, after observing the data $Y$, the posterior distribution is again a mixture distribution (Bernardo and Smith 
\cite{ber1994bt},
O'Hagan and Forster \cite{oha2004bi}, 
Spiegelhalter et al. \cite{spi2004bac})
\begin{equation}
f(\theta \vert Y) = \sum_{i=1}^k \tilde{p}_i f_i(\theta \vert Y)
\end{equation}
with
\begin{enumerate}
\item[(i)]
mixture component distributions being equal to the component-wise posterior
distributions $f_i( \theta \vert Y)$ 
\item[(ii)] 
mixture weights depending on the original weights and the
  marginal likelihoods (prior predictive probabilities) $f_i(Y)$ 
  for each prior distribution as follows 
\begin{equation} 
\label{post::w}
\tilde{p}_i = \frac{p_i f_i(Y)}{\sum_{i=1}^k p_if_i(Y)}
\end{equation}
\begin{equation}
f_i(Y) = \int_{\theta} f(Y \vert \theta) f_i(\theta) d\theta  
\end{equation}
\end{enumerate}
Importantly, (ii) shows that the first concern (non-dynamic (``fixed'') updating) is mistaken. Even though the mixture weights $p_i$ are fixed, the
updating is dynamic: components with higher prior predictive probability $f_i(Y)$ will have higher probability a-posteriori.

\subsubsection{Uncertain mixture weights}
\label{weights::uncertain}
We now consider the extension of (\ref{mixprior::fixed1}) and assume 
uncertain weights $p_i$. 
Using $\pi_i$ instead of $p_i$ and
$\pi=(\pi_1,\ldots,\pi_k)$, $\sum_{i=1}^k
\pi_i=1$, the conditional distribution of $\theta$ given $\pi$
is
\begin{equation}
\label{mixprior:random1}
f(\theta \vert \pi) = \sum_{i=1}^k \pi_i f_i(\theta)
\end{equation}
Since the weights are now uncertain, they require a prior
distribution $f(\pi)$. This implies the joint prior distribution
\begin{equation}
\label{mixprior:random2}
f(\theta,\pi)  = f(\pi) f(\theta \vert \pi)  =  f(\pi)
\sum_{i=1}^k \pi_i f_i(\theta)
\end{equation}
Since the statistical model for $Y$ depends only on $\theta$, the
inference for $\theta$ depends only on 
the marginal prior distribution, which is 
\begin{equation}
f(\theta) = \int_{\pi} f(\theta, \pi) d\pi 
=
\int_{\pi} f(\pi)  \sum_{i=1}^k \pi_i f_i(\theta)  d\pi
=
\sum_{i=1}^k E(\pi_i) f_i(\theta) 
\end{equation}
Thus, the prior distribution $f(\theta)$ depends only on $E(\pi_i)$, which invalidates the second concern.

In summary,
the fixed and uncertain weights versions will give identical inferences for $\theta$ if
$E(\pi_i) = p_i \, (i=1,\ldots,k)$. Thus, thinking that one affects the inference of $\theta$ by accounting for uncertainty of the mixture weights is mistaken.

\subsubsection{An example}
\label{sec::ex}
For illustration, we consider a small proof-of-concept trial in
patients suffering from ankylosing spondylitis, a chronic
inflammatory disease 
(Baeten et al. \cite{bae2013ama}). 
The randomized trial, comparing the monoclonal antibody \emph{secukinumab} to placebo, used a historical data design to leverage placebo data from eight previous trials with a total of 179 placebo patients.  

The authors derived a $\mbox{Beta}(11,32)$ prior for the placebo
response rate $\theta$, using the hierarchical
meta-analytic-predictive approach 
(Neuenschwander et al. \cite{neu2010shi}, 
Schmidli et al. \cite{sch2014rmp}, 
Spiegelhalter et al. \cite{spi2004bac}, 
Viele et al. \cite{vie2014uhc}),
Here, we use a robust mixture version of the original prior to hedge against potential prior-data conflict 
\[
f(\theta) = 0.75\times \mbox{Beta}(11,32)+
0.25\times \mbox{Beta}(1,1)
\]
and compare posterior results to the version with uncertain 
weights.
For illustration, the latter are represented by a $\mbox{Dirichlet}(7.5,2.5)$ prior
distribution for $\pi=(\pi_1,\pi_2)$, equivalent to a
$\mbox{Beta}(7.5,2.5)$ prior for $\pi_1$.
The control data in the actual trial had one responder in the six
placebo patients (r/n=1/6), which are the most likely data under the prior. To illustrate prior-data conflict, we assume data r/n = 4/6, which are in conflict with the original prior.

For fixed weights, the posterior distribution follows from 
Section \ref{weights::fixed}. 
The beta-binomial prior predictive probability for $r/n=4/6$ 
under the original Beta$(11,32)$ prior is $0.043$.
For the uniform prior it is considerably larger $(1/7= 0.143)$, suggesting some prior-data conflict. 
The posterior mixture weights (\ref{post::w}) are $\tilde{p}_1=
(0.75\times0.043)/(0.75\times 0.043+0.25\times 0.143)=0.475$ and 
$\tilde{p}_2=0.525$, considerably different from the prior weights despite the small sample size. Thus, the posterior distribution is
\[
f(\theta \vert r=4,n=6) = 0.475\times \mbox{Beta}(15,34)+0.525\times \mbox{Beta}(5,3)
\]
The posterior mean and 95\%-interval are 
$0.474$ and $(0.202,0.874)$, respectively. In contrast, the posterior mean and 95\%-interval for the non-robust original Beta(11,32) prior would have been problematic given the rather high observed response rate: $0.306$ and $(0.187,0.441)$.

For uncertain weights, MCMC can be used to obtain the posterior
distribution. We have used WinBUGS for the analysis (Appendix \ref{sec::appex1}).
The results for fixed and uncertain weights are summarized in Table
\ref{tab::ex}. Note that the latent variable version provides the
posterior weights for the latent variable: since Z takes values 1 or 2, the mean $1.525$ corresponds to posterior weights 0.475 and 0.525 for components 1 and 2, respectively. The MCMC
analyses of the two versions confirm the theoretical result of Section (\ref{weights::uncertain}), showing identical results obtained with a burn-in of 50'000 and an MCMC sample of 500'000 iterations. 
 


\begin{center}
\begin{table}[h!]
  \caption{Posterior summaries for the response rate for data
    r/n=4/6 for the robust mixture prior of Section \ref{sec::ex} with fixed [1] and
    uncertain ("random") [2] weights, respectively, and for the latent 
    component variable Z taking values 1 and 2.
  }
  
\begin{tabular}{clllll}
       \\[1mm]
	    &	mean	& sd	& 2.5\%	& median & 97.5\% \\[1mm]	
	p[1]&	0.474	&0.203	&0.202	&0.409	&0.874\\
	p[2]&	0.473	&0.203	&0.203	&0.409	&0.873\\
	Z[1]&	1.525	&0.499	&1.0	&2.0	&2.0\\
	Z[2]&	1.525	&0.499	&1.0	&2.0	&2.0\\
\end{tabular}
\label{tab::ex}
\end{table} 
\end{center}

\subsection{Multiple mixture priors}
\label{sec:mixpriorm}
We now consider the case of more than one mixture prior.
That is, for $m$ parameters $\theta_1,\ldots,\theta_m$, the latent variable formulation is
\begin{equation}
f(\theta_j \vert Z_j) = Z_j f_1(\theta_j) + (1-Z_j)f_2(\theta_j) 
\qquad j=1,\ldots,m
\end{equation}
For simplicity, we assume two-component mixtures with the latent variable $Z_j$ following Bernoulli distributions, that is, taking values 1 and 0 (rather than 1 and 2) for mixture component 1 and 
2. 

\subsubsection{Three versions}
Three versions are usually considered for multiple mixture priors, although only two are really relevant. The versions differ in regard to the distribution of the latent variables:
\begin{enumerate}
\item[(1A)]
$Z_j \sim \mbox{Bern}(p_j)$, \quad $p_j$ fixed
\item[(1B)]
$Z_j \sim \mbox{Bern}(\pi_j), \quad \pi_j \sim \mbox{Beta}(a_j,b_j)$
\item[(2)]
$Z_j \sim \mbox{Bern}(\pi), \quad \pi \sim \mbox{Beta}(a,b)$
\end{enumerate}

First, it is easy to see that versions (1A) and (1B) are 
equivalent: the marginal distributions of the latent variables $Z_j$ are $\mbox{Bern}(p_j)$ and 
$\mbox{Bern}(a_j/(a_j+b_j))$, respectively; and, since the latent variables are independent, inferences  for the parameters $\theta_j$ will be equivalent if $p_j=a_j/(a_j+b_j)$. Version (1B) should therefore be avoided: assuming that one introduces additional uncertainty for the mixture weights via prior distributions for each $\pi_j$ is misguided.  

However, version (2) will lead to different inferences for the parameters $\theta_1,\ldots,\theta_m$ even if the fixed mixture weights of version (1A) are assumed identical ($p_1=\ldots,p_m=p$) and equal to the mean of the prior distribution in (2), that is, 
$p=a/(a+b)$. 
While the marginal distributions of $Z_1,\ldots,Z_m$ are still the same for both versions, the multivariate distributions are not. For example (see Appendix \ref{sec::appex2}):
\begin{itemize}
\item
the conditional distribution of $Z_j$ given $Z_k=z$ is 
$ pr(Z_j=1 \vert Z_k=z) = (a+z)(a+b+1)$ for version (2)
and
$ pr(Z_j=1 \vert Z_k=z) = pr(Z_j=1) = a/(a+b)$ for version (1A);
\item
the correlations $cor(Z_j,Z_k)$ are $1/(a+b+1)$ for version (2) and zero for version (1A);
\item
the sum of the latent variables is binomial for version (1A) and Beta-binomial for version (2).
\end{itemize}

\subsubsection{Shrinkage priors}
Version (2), using a common a-priori mixture weight $\pi \sim \mbox{Beta}(a,b)$, is sometimes referred to as a shrinkage prior. Shrinkage follows from the a-priori correlation of the latent variables. The smaller $a$ and $b$, the more shrinkage (similarity) will be introduced for the inference of the $\theta_j$ 
parameters. 
A uniform prior $(a=b=1)$ is typically used, for which the prior correlation $cor(Z_j,Z_k)$ is $1/3$; for the Jeffreys prior, it is 1/2.
Of note, with increasing $a+b$, the inference under version (1A) and (2) will be increasingly similar if $p_1=\ldots=p_m=a/(a+b)$.

Similar to prior specifications in general, how to specify $a$ and $b$ in particular depends on the context. For example, in variable selection problems, the sum of the latent variables (with values in $0,\ldots,m$) has a uniform prior distribution if $a=b=1$; that is, it is uninformative on the number-of-selected-variables scale, which is sometimes used as a justification. However, using large values $a=b$ (leading to the fixed weights version 1A in the limit) leads to equal model probabilities for the $2^m$ models, which is uninformative on the model-probability scale.       Thus, considering informativeness of the prior specification of the latent variables appears illusory.

\section{Summary and recommendations}
\label{sec::discussion}

Mixture prior distributions offer great flexibility to represent prior information, and they are being used in various contexts.
Our focus has been on clarifying two common misunderstandings regarding  fixed and uncertain mixture weights. 

For a single mixture distribution, the two versions yield identical inferences if the fixed component weights $p_1, \ldots,p_k$ are equal to the prior means $E(\pi_j)$. This invalidates two concerns: that updating is non-dynamic (predetermined by the fixed weights), and that uncertainty of the mixture weights is ignored. The sufficiency of fixed weights simplifies the prior specification. For example, for a two-component mixture prior, a single mixture weight needs to be specified, which will depend on the context. In practice, a further 
simplification, which restricts the value to a small number of choices, will often be sufficient; e.g., $p_1 \in  \lbrace 0.9, 0.75, 0.5, 0.25, 0.1 \rbrace$ 
(with corresponding odds 9, 3, 1, 1/3, 1/9).     

The case of mixture priors for multiple parameters $\theta_1,\ldots,\theta_m$ is more complex. Like in the single prior case, the question of non-dynamic updating or ignoring uncertainty remains irrelevant. 
There are two critical questions, however: 
\begin{enumerate}
  \item 
  Is the assumption of equal mixture weights, $Z_j \sim \mbox{Bern}(\pi)$ with a common $\pi$, sensible?
  \item Is a-prior dependency of the latent variables $Z_j$ is sensible? 
\end{enumerate}
The answers depend on the context. If both are answered positively, the shrinkage prior version $\pi \sim \mbox{Beta}(a,b)$ for a two-component prior (or a Dirichlet prior for $k>2$ components) is advised, where smaller values of $a$ and $b$ imply increased shrinkage. Otherwise, fixed weigts $p_1,\ldots,p_m$ should be used.

Finally, the results discussed here should not be generalized. That is, they do not imply that extending prior distributions by additional layers
(distributions on hyperparameters) should be avoided in general.
We have seen that for the case of a single mixture prior with uncertain mixture weights, the marginal prior $f(\theta)$ is equal to the mixture prior with fixed weights.
This equivalence does not hold in general. For example, assume the original prior for a parameter $\theta$ as $N(\mu,\sigma^2)$ with known $\mu$ and $\sigma^2$. Extending this
prior by letting $\sigma^2$ to follow an inverse Gamma
distribution leads to a Student-t prior for $\theta$. All that matters is the marginal prior distribution $f(\theta)$, which is different from the original normal prior. When using such prior extensions, we recommend checking whether the marginal prior $f(\theta)$ reflects prior information appropriately. 

\begin{center}
\begin{table}[h!]
  \caption{Mixture priors in publications}
  \label{PubOverview}
  
\begin{tabular}{lccc}
       \\[1mm]
	    &	 single/multiple &  fixed (F), uncertain (U) & discussion/justification \\ 
Dallow \cite{dal2018bdm})  & single & F &  \\	
Gajewski \cite{gaj2006bss} &  single &  F &  \\
Fridley \cite{fri2010bmm} & single & U &  \\	
Moatti \cite{moa2013moe}  & single &  F &  \\
Mutsvari \cite{mut2016app} & single & F & \\
Walley \cite{wal2015aba} &  single &  F &  \\
Berry \cite{ber2004afm}  & multiple & U & no \\	
Fruehwirth-Schnatter  \cite{fru2010bvs} & multiple & U & no \\
Ley \cite{ley2008ote} & multiple &  U & yes \\
Malsiner-Walli  \cite{mal2011csa} & multiple & U &  yes \\
Papanikos \cite{pap2020bhm} & multiple &  U & yes \\
Vanucci \cite{van2010bmf} & multiple &  U & yes \\
Xia  \cite{xia2011bhm} &  multiple &  U & no \\
Xu \cite{xu2015bvs} &   multiple & U & yes  
\end{tabular}
\end{table} 
\end{center}

We conclude by an overview of publications (Table \ref{PubOverview}), which includes the number of mixture distributions (single vs. multiple), the use of fixed (F) or uncertain (U) mixture weights, and whether a justification for their use is given. For multiple mixture priors, we think a justification or discussion for fixed or uncertain weights is always advised. As we have seen, the fixed version suffices for a single mixture prior, so no justification is needed. 

For an example with multiple mixture priors for variable selecction, see supplementary material (talk by Simon Wandel).

\bibliographystyle{plain}
\bibliography{BenBiblio}

\section{Appendix}
\label{sec::appex}

\subsection{WinBUGS code with fixed and uncertain weights}
\label{sec::appex1}

\begin{verbatim}
model{
# two mixture priors with weight vectors w1 and w2 
# (latent variable version)

pmix1[1] ~ dbeta(beta1[1],beta1[2])
pmix1[2] ~ dbeta(beta2[1],beta2[2])
Z[1] ~ dcat(w1[1:2])
p[1] <- pmix1[Z[1]]

pmix2[1] ~ dbeta(beta1[1],beta1[2])
pmix2[2] ~ dbeta(beta2[1],beta2[2])
w2[1:2] ~ ddirch(w2.dir[1:2])
Z[2] ~ dcat(w2[1:2])
p[2] <- pmix2[Z[2]]

r1 ~ dbin(p[1],n)
r2 ~ dbin(p[2],n)
}

list(
n = 6,
# data replicated for the two analyses
r1 = 4, r2 = 4, 
# parameters of the two mixture components
beta1=c(11,32),
beta2=c(1,1),
# fixed weights w1
# uncertain weights, w2 ~ Dirichlet 
w1 = c(0.75,0.25),
w2.dir = c(7.5,2.5)
)

\end{verbatim}

\subsection{Some results for multiple mixture priors (version 2) }
\label{sec::appex2}

\begin{itemize}
\item
$m$ parameters of interest: $\theta_1,\ldots,\theta_m$.
\item
Each of them follows a mixture distribution: 
$\theta_j \sim F_1$ with probability $\pi$ and
$\theta_j \sim F_2$ with probability $1-\pi$;
the means and variances of $F_1$ and $F_2$ are $\mu_1,\sigma_1^2$ and $\mu_2,\sigma_2^2$, respectively.
\item
The respective latent variables for the two mixture components are the indicators $Z_1,\ldots,Z_m$, which are Bernoulli with parameter $\pi$.
\item
$\pi \in (0,1)$ follows a distribution, 
$ \pi \sim G$, which is typically chosen as Beta(a,b).
\end{itemize}

\begin{eqnarray} 
\theta_j & \sim & Z_j F_1 + (1-Z_j)F_2 \qquad j=1,\ldots,m \\
Z_j & \sim & \mbox{Bern}(\pi) \qquad j=1,\ldots,m \\
\pi & \sim & G
\end{eqnarray}

\subsubsection{Latent variables}
Marginal mean and variance:
\[
E(Z) = E_{\pi}E(Z \vert \pi) = E(\pi)
\]
\begin{eqnarray*}
Var(Z) & = & E_{\pi}Var(Z \vert \pi)+Var_{\pi}E(Z \vert \pi) \\
& = & 
E(\pi(1-\pi)) + Var(\pi)
\\
& = & E(\pi)-(E(\pi))^2 
\\
& = & 
E(\pi)(1-E(\pi))
\end{eqnarray*}

Easier: Z is a Bernoulli variable with parameter $E(\pi)$.

Marginal covariance and correlation:
\begin{eqnarray*}
Cov(Z_1,Z_2) & = & E(Z_1Z_2)-E(Z_1)E(Z_2) \\
& = & E_{\pi}E(Z_1Z_2\vert \pi)-(E(\pi)^2 \\
& = & E(\pi^2)-(E(\pi))^2 \\
& = & Var(\pi)
\end{eqnarray*}

and the correlation is
\[
cor(Z_1,Z_2) =  \frac{Var(\pi)}{E(\pi)(1-E(\pi))}
\]

For $\pi \sim Beta(a,b)$: 
\[
Var(Z) = ab/(a+b)^2
\]
\[
cov(Z_1,Z_2) =  \frac{ab}{(a+b)^2(a+b+1)},\quad cor(Z_1,Z_2) = 1/(a+b+1)
\]

For the standard uninformative priors, $\pi \sim Beta(1,1)$ and
$\pi \sim Beta(0.5,0.5)$, the correlations are 1/3 and 1/2, respectively.

Main results:
\begin{itemize}
\item
The marginal distribution of the latent variables depend only on 
$E(\pi)$. Thus, for $m=1$,
using a prior distribution for $\pi$ or using a fixed mixture weight equal to $E(\pi)$ will lead to identical results for $Z$ and $\theta$.
\item
For $m>1$, however, 
\begin{itemize} 
\item 
using a prior distribution induces prior correlation among the latent variables, whereas for a fixed weight the latent variables are a-priori independent.
\item
the sum of the latent variables follows 
a Beta-binomial distribution if $\pi \sim Beta(a,b)$ and 
a binomial distribution if $\pi$ is fixed.
\item
$f(Z_2=1 \vert Z_1=z_1) = (a+z_1)(a+b+1)$ if $\pi \sim Beta(a,b)$
\item
$f(Z_2=1 \vert Z_1=z_1) = f(Z_2=1) = w$ for fixed weight $w$.
\end{itemize}
\end{itemize}

\subsubsection{Main parameters}

Marginal mean and variance:

\[ E(\theta) = E_Z(\theta \vert Z) = E(Z) \mu_1 + (1-E(Z)) \mu_2
= E(\pi) \mu_1 + (1-E(\pi)) \mu_2
\]

\begin{eqnarray*}
E(\theta) = E_Z(\theta \vert Z) = E(Z) \mu1 +
var(\theta) & = & 
E_zVar(\theta \vert z) + Var_zE(\theta \vert z) \\ 
& = & 
E_z(Z^2\sigma_1^2+(1-Z)^2\sigma_2^2)+Var_z(Z\mu_1+(1-Z)\mu_2) \\ 
& = & 
E(\pi)\sigma_1^2 + E(1-\pi)\sigma_2^2 + (\mu_1-\mu_2)^2Var(Z) \\
& = & 
E(\pi)\sigma_1^2 + E(1-\pi)\sigma_2^2 + (\mu_1-\mu_2)^2
E(\pi)(1-E(\pi))
\end{eqnarray*}

Marginal covariance and correlation:
\begin{eqnarray*}
cov(\theta_1,\theta_2) & = & 
E_{(z_1,z_2)}Cov(\theta_1,\theta_2 \vert Z_1,Z_2)
+Cov_{(z_1,z_2)}E(\theta_1,\theta_2 \vert Z_1,Z_2)
 \\ 
& = &
0 + Cov(Z_1 \mu_1+(1-Z_1)\mu_2,Z_2 \mu_1+(1-Z_2)\mu_2)
 \\ 
& = & 
(\mu_1^2-2\mu_1\mu_2+\mu_2^2)Cov(Z_1,Z_2)
\\
& = & 
(\mu_1-\mu_2)^2 Var(\pi)
\end{eqnarray*}

For $\pi \sim Beta(a,b)$: 
\[
Var(\theta) = \frac{a}{a+b}\sigma_1^2+\frac{b}{a+b}\sigma_2^2+\frac{ab}{(a+b)^2}(\mu_1-\mu_2)^2
\]

\[
cov(\theta_1,\theta_2) =  \frac{ab}{(a+b)^2(a+b+1)}(\mu_1-\mu_2)^2
\]
\[
cor(\theta_1,\theta_2) =  \frac{(\mu_1-\mu_2)^2}{(a+b+1)( (a+b)(\sigma_1^2/b+\sigma_2^2/a)+(\mu_1-\mu_2)^2)}
\]

Main results: 
\begin{itemize}
\item
The marginal distributions of the $\theta_j$ parameters depend only on $E(\pi)$.
\item a prior on $\pi$ induces correlation among the main 
parameters only if $\mu_1 \not= \mu_2$.
\end{itemize}

{\bf Note}. 
The prior results for the $\theta$ parameters have been checked via simulation for the following examples: 
\begin{enumerate}
\item[1.] Mixture of normal distributions
\item[2.] Mixture of a uniform (0,c) and uniform (c,1) distribution; one-sided Bayesian hypothesis tests for $m$ response rates
\item[3.]
Mixture of point null mass and a normal distribution; Bayesian hypothesis tests of a sharp null hypothesis
\end{enumerate}


\end{document}